\documentclass{ioparxiv}

\usepackage{lmodern}
\usepackage{amsmath,amssymb,amsfonts}
\usepackage{algorithmic}
\usepackage{graphicx}
\usepackage{booktabs}
\usepackage{siunitx}
\usepackage{textcomp}
\usepackage[acronym,nohypertypes={acronym}]{glossaries}
\usepackage{xcolor}

\usepackage[
  backend=biber,
  style=numeric,
  sorting=none,
  giveninits=true,
  doi=true,
  url=false,
  isbn=false
]{biblatex}
\usepackage{tikz}
\usetikzlibrary{arrows.meta,positioning,calc,intersections,shapes,angles,patterns,quotes,backgrounds,fit}
\usepackage{circuitikz}
\ctikzset{tripoles/pmos style/emptycircle}
\ctikzset{tripoles/mos style/arrows}

\definecolor{mycolor1}{RGB}{31,119,180}
\definecolor{mycolor2}{RGB}{255,127,14}
\definecolor{mycolor3}{RGB}{44,160,44}
\definecolor{mycolorlight}{RGB}{204,204,204}

\tikzset{
  sum/.style={draw, circle, inner sep=1pt, minimum size=10pt, thick},
  signal/.style={-{Latex[length=3mm, width=2mm]}, thick},
  modsignal/.style={-{Latex[open, length=3mm, width=2mm]}, thick, dashed},
  fblock/.style={draw, thick, minimum height=35, minimum width=40, align=center, rounded corners=5pt},
  sig pic/.pic={
    \draw[thick, black] (-0.5,-0.5) .. controls (0.3,-0.5) and (-0.3,0.5) .. (0.5,0.5);
  },
  heaviside pic/.pic={
      \draw[thick, black]
        (-0.6,-0.4) -- (0, -0.4)   
        -- (0, 0.4)               
        -- (0.6, 0.4);            
    },
  hysteresis pic/.pic={
      \draw[thick, black]
        (-0.6,-0.4) -- (0.3, -0.4)   
        -- (0.3, 0.4)               
        -- (0.6, 0.4) -- (-0.3, 0.4) 
        -- (-0.3, -0.4) -- (-0.6,-0.4);        
    },
  ihysteresis pic/.pic={
      \draw[thick, black]
        (-0.6,0.4) -- (0.3, 0.4)   
        -- (0.3, -0.4)               
        -- (0.6, -0.4) -- (-0.3, -0.4) 
        -- (-0.3, 0.4) -- (-0.6,0.4);        
    },
  sblock/.style={draw, thick, minimum height=55, minimum width=60, align=center, rounded corners=2pt,
  path picture={
  \pic at (path picture bounding box.center) {heaviside pic};
  }
  },
  stblock/.style={draw, thick, minimum height=55, minimum width=60, align=center, rounded corners=2pt,
  path picture={
  \pic at (path picture bounding box.center) {hysteresis pic};
  }
  },
  istblock/.style={draw, thick, minimum height=55, minimum width=60, align=center, rounded corners=2pt,
  path picture={
  \pic at (path picture bounding box.center) {ihysteresis pic};
  }
  }
}

\newacronym{cmos}{CMOS}{Complementary Metal-Oxide-Semiconductor}

\begin{document}

\articletype{Original version submitted on January 30, 2026}

\title{A Fully Tunable Ultra-Low Power Current-Mode Memory Cell in Standard CMOS Technology}

\author{Arthur Fyon$^{1,\dagger}$\orcid{0009-0008-0771-8767}, Loris Mendolia$^{1,\dagger}$\orcid{0000-0003-0270-5736}, Jean-Michel Redouté$^{1}$\orcid{0000-0001-9612-4312}, Alessio Franci$^{1,2}$\orcid{0000-0002-3911-625X} and Guillaume Drion$^{1,*}$\orcid{0000-0002-8076-9500}}

\affil{$^1$Department of Electrical Engineering and Computer Science, University of Li\`ege, Li\`ege, Belgium}

\affil{$^2$WEL-T Department, WEL Research Institute, Wavre, Belgium}

\affil{$^\dagger$These authors contributed equally to this work.}

\affil{$^*$Author to whom any correspondence should be addressed.}

\email{gdrion@uliege.be}


\begin{abstract}
This work introduces a fully tunable, ultra-low power unipolar memory cell inspired by the Schmitt-trigger comparator and designed in \gls{cmos} using only nine transistors. The proposed circuit operates entirely in the current domain and exploits a novel feedback configuration between two interdependent Heaviside-like thresholding elements to produce tunable bistable switching behavior. Its three key parameters---threshold current, hysteresis width, and output gain---are independently tunable via programmable bias currents, enabling flexibility across diverse analog computing applications. Unlike prior Schmitt-trigger designs, it simultaneously achieves current-mode operation, nanowatt-range power consumption, temperature stability, and full tunability, solely using standard MOSFET elements. Schematic-level simulations in a 180 nm \gls{cmos} process confirm robust hysteresis and resilience to device mismatch. Building on this circuit, we develop a complete family of spike-based logic gates using three-level current encoding, where the bistable memory retains the polarity of the last spike on each input indefinitely, enabling asynchronous logic operations without temporal windowing or refresh mechanisms. The same circuit also serves as the primitive for Bistable Memory Recurrent Units in analog neural networks, where the quantized hidden states provide inherent noise immunity. Together, these capabilities position the design as a versatile building block for next-generation neuromorphic processors integrating memory, logic, and recurrent computation.
\end{abstract}

\section{Introduction}

The rapid expansion of edge computing and Internet-of-Things applications has created an urgent demand for energy-efficient hardware capable of performing inference and signal processing directly at the sensor node \cite{edge2}. In conventional von Neumann architectures, the physical separation between memory and processing units leads to significant energy and latency costs associated with data movement, a limitation commonly referred to as the memory wall \cite{sarpeshkar1998analog, sebastian2020memory}. This bottleneck has become increasingly critical as the volume of data generated at the edge continues to grow, motivating the development of computing paradigms where processing occurs directly within or alongside memory \cite{ielmini2018memory, aguirre2024hardware}.

Two main in-memory computing approaches have emerged. Digital implementations based on static random-access memory (SRAM) arrays offer full compatibility with standard \gls{cmos} processes and reliable, deterministic operation \cite{jhang2021challenges, lin2022review, mannocci2026fully}. However, ensuring read stability during computation requires augmented bitcells, and peripheral circuits (converters, sense amplifiers, adder trees) add substantial area and power overhead \cite{noguchi2008best}. Memristive crossbar arrays offer an alternative with exceptional density and massive parallelism, enabling matrix-vector multiplication in a single step \cite{sebastian2020memory, markovic2020physics, joshi2020accurate}. Yet these face fundamental challenges: device-to-device and cycle-to-cycle variability \cite{james2022variability}, programming nonlinearity, sneak path currents requiring selector devices \cite{shi2020research}, and fabrication steps beyond standard \gls{cmos} \cite{aguirre2024hardware}.

These limitations motivate compact bistable elements as alternative primitives combining \gls{cmos} reliability with ultra-low power operation. In neuromorphic architectures, bistable circuits provide noise-immune storage of binary states and quantized synaptic weights \cite{neuro1, neuro2, christensen20222022, khacef2023spike, schuman2022opportunities, rubino2023neuromorphic}. Spike-based logic has also attracted interest, with existing approaches relying on memristive devices with inherent short-term temporal dynamics \cite{gale2019neuromorphic}, conventional logic gates within spiking neural networks \cite{ayuso2023construction, wang2021logicsnn}, or emerging two-dimensional materials \cite{chen2021logic, migliato2020logic}. However, a pure \gls{cmos} implementation capable of performing logic operations on spikes over an arbitrarily long time window without refresh or temporal constraints remains unexplored. More broadly, the growing demand for on-device machine learning at the edge, where power budgets are severely constrained and always-on operation is required, calls for circuit primitives that can support both logic and recurrent computation within nanowatt power envelopes and standard \gls{cmos} technology nodes.

Schmitt triggers are natural candidates for these applications, providing hysteresis and bistable switching \cite{dokic1984cmos, filanovsky2002cmos} while also serving in signal conditioning \cite{wang2002cmos}, threshold detection, and memory cells \cite{kulkarni2007160}. Recent work on the Bistable Memory Recurrent Unit (BMRU) has shown that Schmitt trigger activation functions enable direct analog implementation of recurrent neural networks with quantized hidden states and inherent noise immunity \cite{degeeter2026, brandoit2026cumulative, fyon2026ultra}. Voltage-mode implementations dominate the literature, including inverter-based positive feedback \cite{Vmode1, steyaert1986novel}, stacked inverters \cite{Vmode2}, operational transconductance amplifiers \cite{Vmode3}, current differencing buffered amplifiers \cite{Vmode4}, and current conveyors \cite{Vmode5, Vmode6}. These are often limited by fixed thresholds, reliance on external resistors, or high power. Current-mode designs \cite{wang1988novel, yuan2010high, eldeeb2017low, Imode1, Imode2, Imode3, Imode4, Imode5, Imode6} offer better compatibility with subthreshold operation, particularly in modern technology nodes~\cite{rubino2023neuromorphic}, but typically lack full tunability of threshold and hysteresis width, or consume power above the \SI{}{\micro\watt} range. Table~\ref{tab:comparison} summarizes key features of existing designs.

\begin{table}[t]
\caption{Comparison with existing Schmitt trigger designs}
\label{tab:comparison}
\centering
\begin{tabular}{@{}lccccc@{}}
\toprule
\textbf{Design} & \textbf{Mode} & \textbf{Resistorless} & \textbf{Power range} & \textbf{Tunable}  \\
\midrule
\cite{Vmode1} & Voltage & No (3) & \SIrange{1}{10}{\milli\watt} & Yes \\
\cite{Vmode2} & Voltage & Yes & \SIrange{1}{10}{\micro\watt} & No \\
\cite{Vmode3} & Voltage & No (2) & N/A & Yes \\
\cite{Vmode4} & Voltage & No (3-4) & N/A & No \\
\cite{Vmode5} & Voltage & No (4) & N/A & Partial \\
\cite{Vmode6} & Voltage & No (1) & \SI{100}{\micro\watt} to \SI{1}{\milli\watt} & Yes \\
\cite{Imode1} & Current & Yes & \SIrange{10}{100}{\micro\watt} & Partial \\
\cite{Imode2} & Current & Yes & \SIrange{10}{100}{\micro\watt} & Yes \\
\cite{Imode3} & Current & Yes & \SI{100}{\milli\watt} to \SI{1}{\watt} & Yes \\
\cite{Imode4} & Current & Yes & \SI{100}{\micro\watt} to \SI{1}{\milli\watt} & Yes \\
\cite{Imode5} & Current & No (2) & \SIrange{10}{100}{\micro\watt} & No \\
\cite{Imode6} & Current & Yes & \SIrange{10}{100}{\micro\watt} & Partial  \\
\textbf{Proposed} & Current & Yes & \SIrange{1}{10}{\nano\watt} & Yes  \\
\bottomrule
\end{tabular}
\end{table}

Motivated by the development of ultra-low power spike-based logic intelligence at the edge and by the limitations of existing designs, in this work we present a novel current-mode \gls{cmos} Schmitt trigger simultaneously achieving: (i) \si{\nano\watt}-range power consumption on standard \gls{cmos} technology; (ii) full tunability of thresholds and output current gain; (iii) compact nine-transistor implementation; and (iv) resistorless and capacitorless realization. The proposed circuit is intended for applications in ultra-low-power analog neuromorphic and event-driven systems rather than high-speed digital and mixed-signal logic. We build on this circuit to develop a complete family of spike-based elementary logic gates that operate in the \si{\nano\watt} range to compare the polarities of successive input spikes over an indefinite time window, demonstrating the potential of our tunable current-mode Schmitt trigger as a flexible primitive for asynchronous neuromorphic computation. The limitations of the current design are unipolar operation (positive currents only) and a trade-off between operating frequency, transistor count and size, and power consumption.

\section{Method}
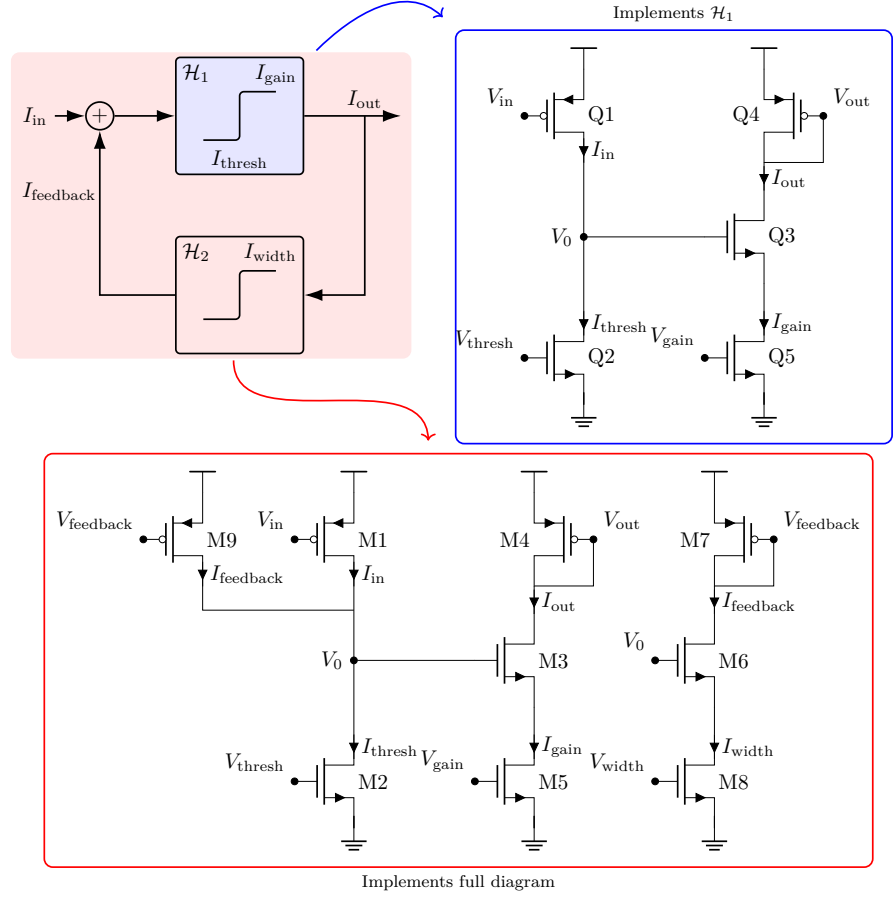
\begin{figure*}[ht!]
    \centering
    \scalebox{0.8}{\pgfdeclarelayer{background}
\pgfsetlayers{background,main}
\begin{circuitikz}[american, transform shape]
    \node[sum] (sum1) {+};
    \node[sblock, right=1. of sum1, label={[anchor=north west]north west:$\mathcal{H}_1$}, label={[anchor=north east]north east:$I_\text{gain}$}, label={[anchor=south]south:$I_\text{thresh}$}] (h1) {};
    \node[sblock, below=of h1, label={[anchor=north west]north west:$\mathcal{H}_2$}, label={[anchor=north east]north east:$I_\text{width}$}] (h2) {};
    
    \node[left=0.5 of sum1] (Iapp) {$I_\text{in}$};
    \coordinate[right=1. of h1, label=above:{$I_\text{out}$}] (Iout) {};
    \coordinate[below left=0.8 and 0.5 of sum1, label=below:{$I_\text{feedback}$}] (Ifb) {};
    
    \draw[signal] (Iapp) -- (sum1) node[pos=1, above] (sig1) {};
    \draw[signal] (sum1) -- (h1) node[midway, above] (sig2) {};
    \draw[signal] (h1) -- (Iout) -- ++(0.6,0) node[midway, right] (sig3) {};
    \draw[signal] (Iout) |- (h2.east) node[pos=0.25, right] (sig4) {};
    \draw[signal] (h2.west) -| (sum1.south) node[pos=0.5, left] (sig5) {};
    \coordinate (leftmargin) at ($(Iout) + (0.7,0)$);

    \draw ($(sum1) +(8,0)$) node[pmos] (q1) {Q1};
    \draw ($(sum1) +(8,-4)$) node[nmos] (q2) {Q2};
    
    \draw (q1.drain) -- (q2.drain) coordinate[midway,circ,label={left:$V_\text{0}$}] (v1);
    \draw (v1) -- ++(2.,0) node[nmos, anchor=G] (q3) {Q3};
    \draw (q2.source) -- ++(0,0.2) node[ground] {};
    \draw (q1.source) node[rground, yscale=-1] {};
    
    \draw ($(q3) +(0,2)$) node[pmos, xscale=-1] (q4) {\ctikzflipx{Q4}};
    \draw (q3.drain) -- (q4.drain);
    \draw (q4.gate) |- (q4.drain);
    \draw (q4.source) node[rground, yscale=-1] {};
    
    \draw ($(q3) -(0,2)$) node[nmos] (q5) {Q5};
    \draw (q3.source) -- (q5.drain);
    \draw (q5.source) -- ++(0,0.2) node[ground] {};
    
    \draw (q2.gate) node[circ,label={above left:$V_\text{thresh}$}] {};
    \draw (q2.drain) to[short, i=$I_\text{thresh}$] ++(0,-0.4);
    
    \draw (q1.gate) node[circ,label={above left:$V_\text{in}$}] {};
    \draw (q1.drain) to[short, i_<=$I_\text{in}$] ++(0,0.4);
    
    \draw (q5.gate) node[circ,label={above left:$V_\text{gain}$}] {};
    \draw (q5.drain) to[short, i=$I_\text{gain}$] ++(0,-0.4);
    
    \draw (q3.drain) to[short, i_<=$I_\text{out}$] ++(0,0.4);
    \draw (q4.gate) node[circ,label={above right:$V_\text{out}$}] {};

    \draw ($(sum1) +(4.2,-7)$) node[pmos] (m1) {M1};
    \draw ($(sum1) +(4.2,-11)$) node[nmos] (m2) {M2};

    \draw (m1.drain) -- (m2.drain) coordinate[midway,circ,label={left:$V_\text{0}$}] (v1_2);
    \draw (v1_2) -- ++(2.,0) node[nmos, anchor=G] (m3) {M3};
    \draw (m2.source) -- ++(0,0.2) node[ground] {};
    \draw (m1.source) node[rground, yscale=-1] {};

    \draw ($(m3) +(0,2)$) node[pmos, xscale=-1] (m4) {\ctikzflipx{M4}};
    \draw (m3.drain) -- (m4.drain);
    \draw (m4.gate) |- (m4.drain);
    \draw (m4.source) node[rground, yscale=-1] {};

    \draw ($(m3) -(0,2)$) node[nmos] (m5) {M5};
    \draw (m3.source) -- (m5.drain);
    \draw (m5.source) -- ++(0,0.2) node[ground] {};

    \draw (m3) ++(2.,0) node[nmos, anchor=G] (m6) {M6};

    \draw ($(m6) +(0,2)$) node[pmos, xscale=-1] (m7) {\ctikzflipx{M7}};
    \draw (m6.drain) -- (m7.drain);
    \draw (m7.gate) |- (m7.drain);
    \draw (m7.source) node[rground, yscale=-1] {};

    \draw ($(m6) -(0,2)$) node[nmos] (m8) {M8};
    \draw (m6.source) -- (m8.drain);
    \draw (m8.source) -- ++(0,0.2) node[ground] {};

    \draw ($(m1) -(2.5,0)$) node[pmos] (m9) {M9};
    \draw (m9.source) node[rground, yscale=-1] {};
    \draw (m9.drain) |- ($(m1.drain) -(0,0.4)$);

    \draw (m2.gate) node[circ,label={above left:$V_\text{thresh}$}] {};
    \draw (m2.drain) to[short, i=$I_\text{thresh}$] ++(0,-0.4);
    
    \draw (m1.gate) node[circ,label={above left:$V_\text{in}$}] {};
    \draw (m1.drain) to[short, i_<=$I_\text{in}$] ++(0,0.4);

    \draw (m5.gate) node[circ,label={above left:$V_\text{gain}$}] {};
    \draw (m5.drain) to[short, i=$I_\text{gain}$] ++(0,-0.4);
    
    \draw (m3.drain) to[short, i_<=$I_\text{out}$] ++(0,0.4);
    \draw (m4.gate) node[circ,label={above right:$V_\text{out}$}] {};

    \draw (m6.gate) node[circ,label={above left:$V_\text{0}$}] {};

    \draw (m6.drain) to[short, i_<=$I_\text{feedback}$] ++(0,0.4);
    \draw (m7.gate) node[circ,label={above right:$V_\text{feedback}$}] {};

    \draw (m8.gate) node[circ,label={above left:$V_\text{width}$}] {};
    \draw (m8.drain) to[short, i=$I_\text{width}$] ++(0,-0.4);

    \draw (m9.gate) node[circ,label={above left:$V_\text{feedback}$}] {};
    \draw (m9.drain) to[short, i_<=$I_\text{feedback}$] ++(0,0.4);

    \coordinate (hbox-nw) at ($(q1) + (-2.,1.3)$);
    \coordinate (hbox-se) at ($(q5) + (2.,-1.3)$);
    
    \node[draw=blue, thick, rounded corners, fit=(hbox-nw)(hbox-se), label=above:{\footnotesize Implements $\mathcal{H}_1$}] (hbox) {};
    
    \coordinate (fbox-nw) at ($(m9) + (-2.5,1.3)$);
    \coordinate (fbox-se) at ($(m8) + (2.5,-1.3)$);
    
    \node[draw=red, thick, rounded corners, fit=(fbox-nw)(fbox-se), label=below:{\footnotesize Implements full diagram}] (fbox) {};

    \draw[->, blue, thick] ($(h1.north east) + (0.2,0.1)$) to[out=45, in=135] ($(hbox.north west) + (-0.2,0.2)$);
    
    \draw[->, red, thick] ($(h2.south) + (-0.1,-0.1)$) to[out=-90, in=90] ($(fbox.north) + (-0.5,0.2)$);

    \begin{pgfonlayer}{background} 
      \node[
        fill=red!10,
        rounded corners,
        inner sep=2pt,
        fit=(sum1)(h1)(h2)(Iapp)(Iout)(Ifb)(sig1)(sig2)(sig3)(sig4)(sig5)(leftmargin)
      ] (redbox) {};

      \node[
        fill=blue!10,
        rounded corners,
        inner sep=0pt,
        fit=(h1)
      ] {}; 
    \end{pgfonlayer}
\end{circuitikz}}
    \caption{Conceptual architecture of the proposed dual-Heaviside feedback Schmitt trigger, along with the electronic schematics of the individual Heaviside circuit and the full Schmitt trigger.}
    \label{fig:heaviside}
\end{figure*}

The proposed circuit is based on a dual-Heaviside feedback structure, conceptually illustrated in Figure~\ref{fig:heaviside} (top left panel). It consists of two interdependent nonlinear thresholding elements:
\begin{itemize}
    \item A primary Heaviside element that activates when the input current $I_{\text{in}}$ exceeds a threshold $I_{\text{thresh}}$ (the high threshold of the Schmitt trigger), producing an output current $I_{\text{gain}}$ (the high output value).
    \item A secondary Heaviside element in feedback, triggered by the output of the first stage. It activates immediately to inject a feedback current $I_{\text{width}}$ at the input of the circuit.
\end{itemize}

The hysteresis behavior arises naturally from the feedback current $I_{\text{width}}$, which lowers the effective input threshold to $I_{\text{thresh}} - I_{\text{width}}$, thus enabling bistability. The three control parameters $I_{\text{thresh}}$, $I_{\text{gain}}$, and $I_{\text{width}}$ are all independently adjustable via bias voltages, allowing straightforward tuning. Because the circuit operates only with positive currents, the constraint $I_{\text{thresh}} > I_{\text{width}}$ must be met to preserve bistability, ensuring that the positive feedback does not override the comparator threshold.

The top right panel (blue) of Figure~\ref{fig:heaviside} shows a minimal \gls{cmos} circuit that implements a fully tunable unipolar Heaviside function ($\mathcal{H}_1$) in current mode. The input current is provided through a voltage $V_\text{in}$ at the output of a PMOS current mirror (Q1), while the threshold current $I_{\text{thresh}}$ is set through a voltage $V_\text{thresh}$ by an NMOS transistor (Q2). Together, Q1 and Q2 form a switching current comparator. If $I_{\text{in}} < I_{\text{thresh}}$, then $V_0 \approx 0$ and Q3 is off, resulting in a zero output current $I_{\text{out}}$. As soon as $I_{\text{in}} > I_{\text{thresh}}$, $V_0$ increases to the supply voltage and Q3 turns on, allowing current to flow in the output branch. At that point, the current flowing through this branch corresponds to the bias current $I_{\text{gain}}$, also set through a voltage $V_\text{gain}$ by an NMOS transistor (Q5). The output current flows through the diode-connected transistor Q4 and can be mirrored to other stages via a PMOS transistor driven by $V_\text{out}$.

The bottom panel (red) of Figure~\ref{fig:heaviside} represents the feedback interconnection of the two Heaviside functions, forming a current-mode unipolar Schmitt trigger capable of operating with \si{\nano\watt}-level supply power. $\mathcal{H}_1$ (Q1--Q5) is implemented through transistors M1--M5. For $\mathcal{H}_2$ to switch concurrently with $\mathcal{H}_1$, they share the same switching voltage $V_0$, removing the need for a separate comparator branch (like M1--M2). This means that if $I_{\text{in}} < I_{\text{thresh}}$, M6 is off and $I_\text{feedback} = I_\text{out} = 0$. Once $I_{\text{in}} > I_{\text{thresh}}$, M6 turns on and $I_\text{feedback}$ reaches $I_\text{width}$, the bias current of the feedback branch set through $V_\text{width}$ by an NMOS transistor (M8). This feedback current is mirrored through a PMOS current mirror (M7 and M9), and injected back into the comparator branch of $\mathcal{H}_1$ (M1, M2, and M9), thereby increasing the input current by $I_\text{feedback}$.

The resulting Schmitt trigger characteristics can be expressed as:
\begin{align}
    I_{\text{th,high}} &= I_{\text{thresh}} \\
    I_{\text{th,low}} &= I_{\text{thresh}} - I_{\text{width}} \\
    \Delta I_{\text{hyst}} &= I_{\text{width}}
\end{align}
where $I_{\text{th,high}}$ and $I_{\text{th,low}}$ denote the upper and lower switching thresholds, respectively, and $\Delta I_{\text{hyst}}$ is the hysteresis width. The output current switches between approximately zero and $I_{\text{gain}}$. Beyond the listed metrics in Table~\ref{tab:comparison}, the proposed circuit achieves its performance with only nine transistors, hence improving on transistor count as compared to previous designs.

The circuit was simulated using the X-FAB \SI{180}{\nano\meter} \gls{cmos} process. All NMOS transistors were sized with a width of \SI{5}{\micro\meter} and a length of \SI{5}{\micro\meter}, while PMOS transistors were sized with a width of \SI{5.5}{\micro\meter} and a length of \SI{5}{\micro\meter}. This conservative sizing reduces mismatch when properly laid out and mitigates undesired short-channel effects. The exceptions are transistors M3 and M6, which operate primarily as switches and were sized with a width of \SI{2}{\micro\meter} and minimal length. While the relatively large transistor sizes mitigate mismatch, they also significantly increase parasitic capacitance, limiting the bandwidth to approximately \SI{1}{\kilo\hertz}. This bandwidth is well-suited to bio-medical and human-in-the-loop edge applications. In such contexts, ultra-low power consumption takes precedence over high-speed operation.

\section{Results}

All schematic-level simulations were conducted using SPICE with standard X-FAB \SI{180}{\nano\meter} models, with a supply voltage of \SI{1.8}{V}. Unless otherwise stated, the following baseline parameters were used: $I_\text{gain} = \SI{486}{\pico\ampere}$, $I_\text{thresh} = \SI{368}{\pico\ampere}$, and $I_\text{width} = \SI{216}{\pico\ampere}$. These values yield switching thresholds of \SI{150}{\pico\ampere} and \SI{350}{\pico\ampere} with a high output state of \SI{500}{\pico\ampere}. These values represent a representative operating point chosen to demonstrate the circuit functionality. In practice, bias currents are generated from a stable reference current via standard current mirrors, and the linear tunability demonstrated in subsequent sections allows straightforward adjustment to application-specific requirements.

The following analyses evaluate transient behavior, DC characteristics, robustness to mismatch and temperature, tunability, and spike-based logic operations.

\subsection{Transient Response and DC Characteristics}

\begin{figure*}[ht!]
    \centering
    \includegraphics[width=1\linewidth]{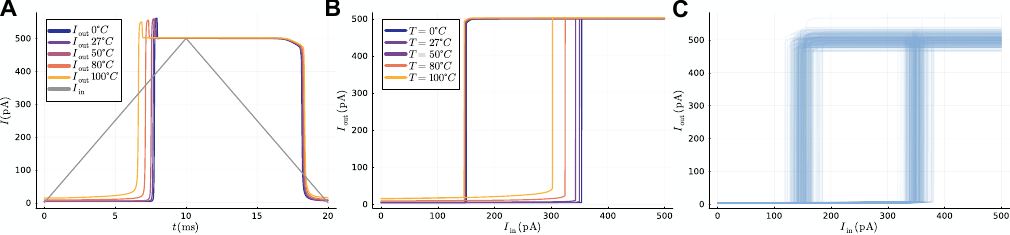}
    \caption{\textbf{A.} Transient simulation of the proposed Schmitt trigger under triangular input current for different operating temperatures. \textbf{B.} DC sweep demonstrating hysteretic behavior for different operating temperatures. \textbf{C.} Monte Carlo analysis of the proposed Schmitt trigger in DC input sweep analysis with $3\sigma$ process variation at room temperature. Baseline parameters: $I_\text{gain} = \SI{486}{\pico\ampere}$, $I_\text{thresh} = \SI{368}{\pico\ampere}$, and $I_\text{width} = \SI{216}{\pico\ampere}$.}
    \label{fig:transient}
\end{figure*}

To evaluate hysteresis, the input current $I_{\text{in}}$ was swept linearly from 0 to \SI{500}{\pico\ampere} and then back to 0 using a triangular waveform. As shown in Figure~\ref{fig:transient}A for different operating temperatures, the circuit exhibits clear bistable behavior with sharp transitions. The switching points align with the expected levels of $I_{\text{thresh}} - I_{\text{width}}$ and $I_{\text{thresh}}$, and the high-state output current closely follows $I_\text{gain}$. A small ($\sim$\SI{10}{\percent}) overshoot is observed at transitions, attributed to the positive feedback inherent in the Heaviside interconnection.

DC simulations confirm the existence of two stable states and well-defined hysteresis (Figure~\ref{fig:transient}B). Both the threshold currents and the high-state output level are independently tunable through their respective bias currents. In both cases, temperature mainly affects the upper switching point of the hysteresis, which is determined by the M1--M2 comparator. However, since the circuit operates in the ultra-low power regime, typical operating conditions remain well below \SI{80}{\celsius}, ensuring robust operation over temperature.

\subsection{Mismatch Analysis}
Monte Carlo simulations were performed under a $3\sigma$ mismatch variation to evaluate the robustness of the design to device mismatch. The circuit consistently maintained correct bistable switching behavior, with only minor deviations, typically a few tens of \si{\pico\ampere}, in the threshold currents, hysteresis width, and high output level (Figure~\ref{fig:transient}C). These variations remain well within functional margins and do not compromise the qualitative operation of the Schmitt trigger.

\subsection{Tuning Properties}
We systematically swept each bias current to assess its effect on the output behavior. Results confirm that each parameter ($I_\text{gain}$, $I_\text{thresh}$, and $I_\text{width}$) controls its corresponding function (output level, threshold, and hysteresis width) in a linear and independent manner for different operating temperatures.

\begin{figure*}[ht!]
    \centering
    \includegraphics[width=1\linewidth]{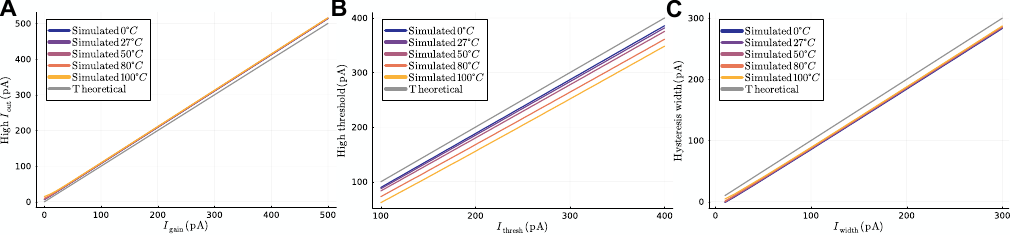}
    \caption{\textbf{A.} $I_\text{gain}$ DC sweep analysis from 0 to \SI{500}{\pico\ampere} of the proposed Schmitt trigger for different operating temperatures. \textbf{B.} $I_\text{thresh}$ DC sweep analysis from \SI{100}{\pico\ampere} to \SI{400}{\pico\ampere} of the proposed Schmitt trigger for different operating temperatures with $I_\text{width} = \SI{50}{\pico\ampere}$. \textbf{C.} $I_\text{width}$ DC sweep analysis from \SI{10}{\pico\ampere} to \SI{300}{\pico\ampere} of the proposed Schmitt trigger for different operating temperatures. Baseline parameters: $I_\text{gain} = \SI{486}{\pico\ampere}$, $I_\text{thresh} = \SI{368}{\pico\ampere}$, and $I_\text{width} = \SI{216}{\pico\ampere}$.}
    \label{fig:params}
\end{figure*}

Figure~\ref{fig:params}A shows the effect of sweeping $I_\text{gain}$ from 0 to \SI{500}{\pico\ampere}. As expected, the high output level increases linearly, although a small offset is present due to subthreshold leakage currents (on the order of a few \si{\pico\ampere}). As a result, the relative error between $I_\text{gain}$ and the actual output current decreases with increasing current, reaching a low \SI{2.8}{\percent} at \SI{500}{\pico\ampere} at room temperature.

Similar results were obtained for threshold and hysteresis width. For $I_\text{thresh}$, the sweep range was limited to \SI{100}{\pico\ampere}--\SI{400}{\pico\ampere} to maintain bistability and avoid approaching either 0 or $I_\text{gain}$. For $I_\text{width}$, the range was \SI{10}{\pico\ampere}--\SI{300}{\pico\ampere} to ensure $I_\text{width} < I_\text{thresh}$ throughout. Figures~\ref{fig:params}B and~\ref{fig:params}C show that the corresponding thresholds and hysteresis widths also follow a linear trend with minor offsets. The relative errors at room temperature were measured at \SI{4.5}{\percent} and \SI{5.25}{\percent}, respectively, for the highest tested currents. These analyses confirm that temperature mainly affects $I_\text{thresh}$, while lowering $I_\text{width}$ to \SI{50}{\pico\ampere} did not change this effect compared with Figure~\ref{fig:transient}B.

\subsection{Spike-Based Logic Operations}
The proposed Schmitt trigger can implement spike-based logic operations by exploiting its hysteretic response and bistable memory. The unipolar constraint naturally accommodates a three-level current encoding scheme: \SI{0}{\pico\ampere} represents logic 0 (negative spike), \SI{500}{\pico\ampere} represents logic 1 (positive spike), and the intermediate resting state at \SI{250}{\pico\ampere} indicates that no spike has yet been received, corresponding to an undefined logic value. This encoding enables bidirectional signaling within the strictly positive current domain.

The key property enabling spike-based logic operations is that the bistable nature of the circuit provides persistent memory of the polarity of the most recent received spike: once an input spikes positively or negatively, that state is retained until the next spike arrives. The logic operation is therefore determined solely by the polarity of the last spike received on each of the logic gate inputs, irrespective of the time elapsed since those spikes occurred. When configured with appropriate bias currents and input summation, the circuit responds selectively to specific polarity combinations: an AND gate activates only when both inputs last spiked positively; an OR gate activates when either input last spiked positively; NAND and NOR gates invert these responses; and an XOR gate activates only when the two inputs differ in their last spike polarity. This polarity-based encoding mirrors neuromorphic computing principles~\cite{ayuso2023construction, gale2019neuromorphic, wang2021logicsnn, chen2021logic, khacef2023spike}, while the persistent memory eliminates the need for temporal windowing or refresh circuits, a capability not demonstrated by prior memristive or conventional \gls{cmos} spike-based approaches.

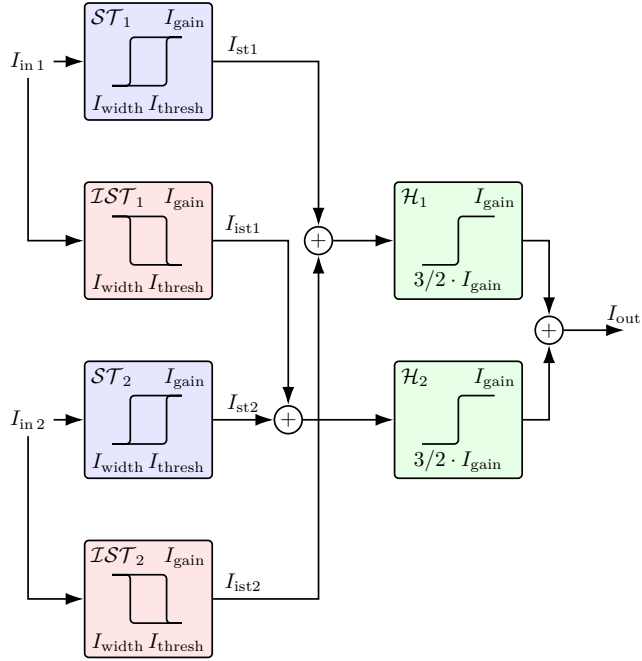
\begin{figure*}[ht!]
    \centering
    \scalebox{0.8}{\pgfdeclarelayer{background}
\pgfsetlayers{background,main}
\begin{circuitikz}[american, transform shape]
    \node[stblock, label={[anchor=north west]north west:$\mathcal{ST}_1$}, label={[anchor=north east]north east:$I_\text{gain}$}, label={[anchor=south]south:$I_\text{width}\, I_\text{thresh}$}] (st1) {};
    \node[istblock, below=of st1, label={[anchor=north west]north west:$\mathcal{IST}_1$}, label={[anchor=north east]north east:$I_\text{gain}$}, label={[anchor=south]south:$I_\text{width}\, I_\text{thresh}$}] (ist1) {};
    
    \node[left=0.5 of st1] (Iapp1) {$I_\text{in\,1}$};
    \coordinate[right=0.5 of st1, label=above:{$I_{\text{st}1}$}] (Iout1pos) {};
    \coordinate[right=0.5 of ist1, label=above:{$I_{\text{ist}1}$}] (Iout1neg) {};
    
    \draw[signal] (Iapp1) -- (st1) node[pos=1, above] (sig1) {};
    \draw[signal] (Iapp1) |- (ist1) node[pos=0.5, left] (sig4) {};

    \node[stblock, below=of ist1, label={[anchor=north west]north west:$\mathcal{ST}_2$}, label={[anchor=north east]north east:$I_\text{gain}$}, label={[anchor=south]south:$I_\text{width}\, I_\text{thresh}$}] (st2) {};
    \node[istblock, below=of st2, label={[anchor=north west]north west:$\mathcal{IST}_2$}, label={[anchor=north east]north east:$I_\text{gain}$}, label={[anchor=south]south:$I_\text{width}\, I_\text{thresh}$}] (ist2) {};
    
    \node[left=0.5 of st2] (Iapp2) {$I_\text{in\,2}$};
    \coordinate[right=0.5 of st2, label=above:{$I_{\text{st}2}$}] (Iout2pos) {};
    \coordinate[right=0.5 of ist2, label=above:{$I_{\text{ist}2}$}] (Iout2neg) {};
    
    \draw[signal] (Iapp2) -- (st2) node[pos=1, above] (sig5) {};
    \draw[signal] (Iapp2) |- (ist2) node[pos=0.5, left] (sig8) {};

    \node[sum, right=1.5 of ist1] (sum1) {+};
    \draw[signal] (st1) -| (sum1) node[pos=0.5, left] (sig10) {};
    \draw[signal] (ist2) -| (sum1) node[pos=0.5, left] (sig11) {};

    \node[sum, right=1. of st2] (sum2) {+};
    \draw[signal] (st2) -- (sum2) node[pos=0.5, left] (sig12) {};
    \draw[signal] (ist1) -| (sum2) node[pos=0.5, left] (sig13) {};

    \node[sblock, right=1. of sum1, label={[anchor=north west]north west:$\mathcal{H}_1$}, label={[anchor=north east]north east:$I_\text{gain}$}, label={[anchor=south]south:$3/2\cdot I_\text{gain}$}] (thresh1) {};
    \draw[signal] (sum1) -- (thresh1) node[pos=0.5, left] (sig14) {};
    
    \node[sblock, right=1.5 of sum2, label={[anchor=north west]north west:$\mathcal{H}_2$}, label={[anchor=north east]north east:$I_\text{gain}$}, label={[anchor=south]south:$3/2\cdot I_\text{gain}$}] (thresh2) {};
    \draw[signal] (sum2) -- (thresh2) node[pos=0.5, left] (sig15) {};

    \node[sum] (sum3) at ($(thresh1)!0.5!(thresh2) + (1.5,0)$) {+};
    \draw[signal] (thresh1) -| (sum3) node[pos=0.5, left] (sig16) {};
    \draw[signal] (thresh2) -| (sum3) node[pos=0.5, left] (sig17) {};
    \coordinate[right=1. of sum3, label=above:{$I_{\text{out}}$}] (Iout) {};
    \draw[signal] (sum3) -- (Iout) node[pos=1, above] (sig18) {};

    \begin{pgfonlayer}{background} 
      \node[
        fill=blue!10,
        rounded corners,
        inner sep=0pt,
        fit=(st1)
      ] {}; 

      \node[
        fill=red!10,
        rounded corners,
        inner sep=0pt,
        fit=(ist1)
      ] {}; 

      \node[
        fill=blue!10,
        rounded corners,
        inner sep=0pt,
        fit=(st2)
      ] {}; 

      \node[
        fill=red!10,
        rounded corners,
        inner sep=0pt,
        fit=(ist2)
      ] {}; 

      \node[
        fill=green!10,
        rounded corners,
        inner sep=0pt,
        fit=(thresh1)
      ] {}; 

      \node[
        fill=green!10,
        rounded corners,
        inner sep=0pt,
        fit=(thresh2)
      ] {}; 
    \end{pgfonlayer}
\end{circuitikz}}
    \caption{Circuit implementation of the XOR logic gate using the proposed Schmitt trigger. Each input is processed by both a standard and an inverted Schmitt trigger to detect spike polarity. The outputs are summed and thresholded to produce the XOR operation. The other four logic gates (AND, OR, NAND, NOR) follow analogous configurations with different summation and threshold arrangements. For the AND gate, $I_{\text{st}1}$ and $I_{\text{st}2}$ (both standard outputs) are summed; for the NOR gate, $I_{\text{ist}1}$ and $I_{\text{ist}2}$ (both inverted outputs) are summed; the OR and NAND gates use complementary summation configurations with appropriate threshold inversions.}
    \label{fig:logic_gates}
\end{figure*}

Figure~\ref{fig:logic_gates} shows the circuit implementation of the XOR gate, from which the other four logic gates are readily derived. Each input is fed into both a Schmitt trigger and an inverted Schmitt trigger to detect spike polarity. These outputs are then summed depending on the desired logic operation. For instance, the NOR gate sums both inverted Schmitt trigger outputs and passes them through a thresholding element at \SI{750}{\pico\ampere} that outputs high when both inverted outputs are high, indicating that the last spikes on both input branches were negative. Similarly, the AND gate sums both non-inverted outputs with a threshold detecting when both last spikes were positive, while OR and NAND gates use analogous configurations. The XOR gate exploits asymmetric hysteresis to distinguish between matched polarities (both positive or both negative) and mismatched polarities (one positive, one negative).

\begin{figure*}[ht!]
    \centering
    \includegraphics[width=1\linewidth]{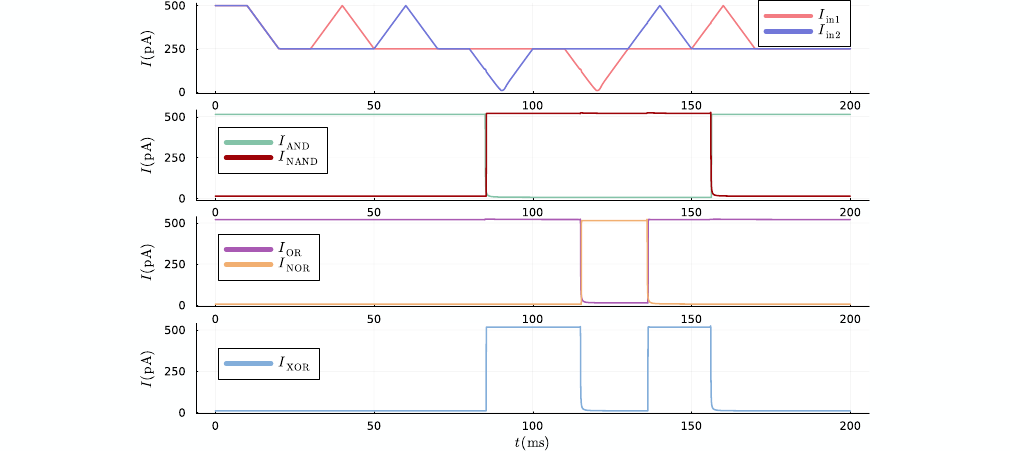}
    \caption{Simulation results of all five spike-based logic gates (AND, NAND, OR, NOR, XOR) for different input spike polarity combinations (top). The three-level current encoding (\SI{0}{\pico\ampere} = logic 0 / negative spike, \SI{250}{\pico\ampere} = undefined / no spike received, \SI{500}{\pico\ampere} = logic 1 / positive spike) enables bidirectional logic operations while maintaining the unipolar current constraint. Each gate correctly responds to the polarity of the last spike on each input, with outputs persisting indefinitely until new spikes arrive. The power consumption of these gates remains in the tens of \si{\nano\watt} range.}
    \label{fig:logic_simus}
\end{figure*}

Figure~\ref{fig:logic_simus} presents simulation results for all five logic gates under different spike polarity combinations. The simulations confirm that the three-level encoding preserves the strictly positive current constraint while enabling full logic functionality with persistent memory.

These spike-based logic gates inherit the nanowatt-range power consumption of the base Schmitt trigger, making them suitable for massively parallel architectures. The hysteresis provides inherent noise immunity, as spurious fluctuations do not trigger state changes. The gates can be cascaded to build more complex computational primitives such as half-adders, full-adders, and sequential logic elements. Unlike purely digital implementations, these analog gates operate entirely within the positive current domain and preserve spike polarity for arbitrarily long time intervals, enabling asynchronous event-driven processing where logic operations occur only when new spikes arrive.

The ability to perform logic operations directly in the current domain based solely on spike polarity is particularly valuable for asynchronous neuromorphic processors and event-driven edge computing~\cite{christensen20222022}. The persistent memory property enables ultra-sparse computation where power is consumed only during spike transitions. Combined with the BMRU recurrent processing capability described in the introduction, these logic gates enable compact neuromorphic processors integrating memory, computation, and decision-making in a unified current-mode architecture.

\section{Discussion}
The proposed Schmitt trigger and its derived logic gates address a gap in the neuromorphic computing landscape: the need for a pure \gls{cmos} primitive that combines bistable memory with ultra-low power operation, without the variability and fabrication complexity of memristive devices or the area overhead of augmented SRAM cells.

The observed mismatches in the Monte Carlo analysis arise primarily from variations in current mirrors and biasing transistors. Their impact could be further reduced by increasing device dimensions, though this would increase the silicon footprint, a trade-off dependent on the intended application and area constraints. Conversely, decreasing transistor dimensions risks compromising the reliability of the hysteresis by causing undesirable overlaps between the low and high switching thresholds, which explains the relatively large transistor sizes selected for our simulations.

The predictable offsets observed in the tunability analysis do not impair functionality. The relationship between control currents and actual switching parameters remains monotonic and approximately linear despite minor offsets due to subthreshold leakage. These deviations can be compensated by simple calibration (\textit{e.g.}, lookup table or linear correction), allowing accurate control of thresholds and hysteresis without additional circuitry.

For the spike-based logic gates, the three-level current encoding scheme introduces additional design considerations. The threshold placement for each gate must account for process variations to ensure correct discrimination between polarity combinations. However, the same tunability that enables flexible Schmitt trigger configuration also allows adjustment of the summing and thresholding stages, providing robustness against variability. The persistent memory property, where spike polarity is retained until explicitly updated, eliminates timing constraints that complicate conventional spike-based logic implementations, but requires that input spikes be sufficiently separated to avoid overlap during the transient response.

The tunability and consistent operation of this circuit allow it to retain functionality under higher bias currents, in the order of \si{\nano\ampere} or \si{\micro\ampere}. This flexibility enables use in both power-critical scenarios and applications where signal integrity and low noise are paramount. Compared with memristive approaches that offer higher density but suffer from variability and nonstandard manufacturing processes, and SRAM-based approaches that provide reliability but at higher area and power cost, the proposed design occupies a distinct niche: pure \gls{cmos} implementation with nanowatt operation and persistent state retention.

\section{Conclusion}
This work introduces a compact, fully tunable, and ultra-low power Schmitt trigger operating in the current domain, implemented with nine transistors in standard \gls{cmos} technology in the subthreshold regime. Unlike previous designs, the proposed solution operates without resistors or capacitors, enabling monolithic integration while offering independent configurability of threshold current, hysteresis width, and output gain. The bistable behavior, achieved through a novel dual-Heaviside feedback architecture, yields robust transitions within \si{\nano\watt} power budgets. Schematic-level simulations confirm correct hysteretic behavior, independent tunability of circuit properties, and resilience to simulated device mismatch and temperature variations.

Building on this circuit, we developed a complete family of spike-based logic gates using a three-level current encoding scheme that represents positive spikes, negative spikes, and a resting state within the strictly positive current domain. The key innovation is that the bistable memory retains the polarity of the last spike on each input indefinitely, enabling logic operations that compare spike polarities without temporal windowing or refresh mechanisms, a capability not reported in prior memristive spike-based logic implementations, which exhibit inherent temporal dynamics, or by conventional digital implementations. All fundamental Boolean operations (AND, OR, NAND, NOR, XOR) were demonstrated through SPICE simulations, with each gate inheriting the nanowatt-range power consumption and noise immunity of the base Schmitt trigger. These gates can be cascaded to construct more complex computational primitives, enabling complete neuromorphic processors operating in the current domain.

Beyond spike-based logic, the proposed Schmitt trigger directly implements the BMRU, a specific RNN whose input-output relationship corresponds one-to-one with the Schmitt trigger transfer function~\cite{degeeter2026, brandoit2026cumulative}. Because the hidden states are quantized, noise and device variations do not accumulate across time steps, providing inherent robustness without explicit error correction~\cite{joshi2020accurate, sebastian2020memory, fyon2026ultra}. This circuit has been employed in a hardware-software co-design demonstration achieving over 90\% accuracy on keyword spotting tasks at approximately \SI{100}{\nano\watt} power consumption~\cite{fyon2026ultra}, suggesting at least three orders of magnitude improvement over TinyML implementations achieving comparable accuracy~\cite{mazumder2022fast, banbury2021mlperf, lin2020mcunet, zhang2017hello}.

The dual functionality demonstrated in this work, spike-based logic and recurrent neural network primitive, positions the proposed Schmitt trigger as a versatile building block for next-generation neuromorphic hardware. The same circuit can serve as a memory element in recurrent layers and as a logic gate in combinatorial processing stages, enabling unified current-mode architectures where memory, computation, and decision-making are tightly integrated. The persistent polarity memory provides both noise immunity and ultra-sparse asynchronous operation, where logic states persist without dynamic power consumption between spike events. Future work will focus on layout implementation, post-layout simulation, and experimental silicon validation to characterize the impact of layout parasitics on bandwidth and matching.

%
%

\ack{The technology described in this paper is the subject of a U.S. Provisional Patent Application No. 63/897,109 and European Patent Application No. EP26175243.0, entitled “Current-mode switching circuits and analog computational systems”.}

\funding{Arthur Fyon is a Postdoctoral Researcher of the Fonds de la Recherche Scientifique -- FNRS, supported by grant ASP-REN 40024838. Loris Mendolia is a FRIA Grantee of the Fonds de la Recherche Scientifique -- FNRS, supported by grant FRIA-B2 40029909. This work was supported by the Belgian Government through the Federal Public Service Policy and Support, under grant NEMODEI2.}

\roles{\begin{itemize}
    \item Arthur Fyon: Conceptualization (lead), Data curation, Formal analysis (lead), Investigation (lead), Methodology (equal), Software, Visualization (lead), Writing - original draft (lead), Writing - review and editing (lead).
    \item Loris Mendolia: Conceptualization (lead), Data curation, Formal analysis (lead), Investigation (lead), Methodology (equal), Software, Visualization (lead), Writing - original draft (lead), Writing - review and editing (lead).
    \item Jean-Michel Redouté: Conceptualization (supporting), Formal analysis (supporting), Funding acquisition (equal), Investigation (supporting), Methodology (equal), Project administration (supporting), Resources (equal), Supervision (supporting), Writing - original draft (supporting), Writing - review and editing (supporting).
    \item Alessio Franci: Conceptualization (lead), Formal analysis (supporting), Funding acquisition (equal), Investigation (supporting), Methodology (equal), Project administration (equal), Resources (equal), Supervision (equal), Writing - original draft (supporting), Writing - review and editing (supporting).
    \item Guillaume Drion: Conceptualization (lead), Formal analysis (supporting), Funding acquisition (equal), Investigation (supporting), Methodology (equal), Project administration (equal), Resources (equal), Supervision (equal), Writing - original draft (supporting), Writing - review and editing (supporting).
\end{itemize}}

\data{All data that support the findings of this study are included within the article. Cadence Virtuoso design files (schematics, symbols, and Maestro configurations) are not included in the public repository due to software licensing constraints and version compatibility issues. These files are available upon reasonable request to the corresponding author for researchers with access to Cadence tools.}


\suppdata{Transistors dimensions optimized for mismatch used in all simulations presented in this work are reported in Table~\ref{tab:sizing}.

\begin{table}[htbp]
\centering
\caption{Ultra-low power Schmitt trigger circuit component dimensions.}
\begin{tabular}{|c|c|c|c|c|c|c|c|c|c|}
\hline
Transistors      & Q/M1 \& Q/M4 \& M7 \& M9 & Q/M2 \& Q/M5 \& M8 & Q/M3 \& M6  \\
\hline
$W \times L$ (\si{\micro\meter}) & $5.5 \times 5$ & $5 \times 5$ & $2 \times 0.18$  \\
\hline
\end{tabular}
\label{tab:sizing}
\end{table}
}

\printbibliography

\end{document}